\documentstyle[amssymb,twocolumn,eqsecnum,aps]{revtex}

\begin{document}
\draft
\title{Nonlinear regime of the current instability\\
in a ballistic field effect transistor}
\author{M. V. Cheremisin}
\address{A.F.Ioffe Physical-Technical Institute, St. Petersburg, Russia}
\maketitle

\begin{abstract}
The strongly nonlinear regime of Dyakonov-Shur instability is studied
analytically and by computer simulations. The instability leads to
high-amplitude stationary oscillations caused by shock wave formation in the
FET channel. The numerically found shape and speed of the shock waves are
well described in terms of the conventional hydrodynamics. The amplitude of
stationary oscillations is determined.
\end{abstract}


\section{Introduction}

Dyakonov and Shur$^{\cite{Dyakonov}}$have described a new mechanism of
plasma wave generation in an ultrashort-channel (ballistic) Field Effect
Transistor. It was shown that 2D-electrons in a BFET can be described by
equations analogous to the hydrodynamic equations for shallow water, the
plasma waves being similar to shallow water waves. Therefore, various
hydrodynamic phenomena observed in shallow water may occur in the
2D-electron fluid. In Ref.$\cite{Dyakonov}$, it was demonstrated that the
non-symmetric boundary conditions with fixed source potential and drain
current lead to instability of the stationary state with dc current. The FET
channel is a resonator cavity, with plasma wave generation mechanism similar
to that known for self-excitation of jets and organ pipes. Both the
eigenfrequencies and the instability increment were found for
dissipationless electrons. The plasma wave generation in short channel FETs
is very important for using these devices as high-power sources in the
terahertz frequency range. The experimental results$^{\cite{Weikle},\cite{Lu}%
}$ found for the response of GaAs HEMT in the detector mode agrees with
theoretical predictions.$^{\cite{Dyakonov2}}$ Then, the recent experiments$^{%
\cite{Burke}}$ concerning GHz plasma waves in 2DEG make an observation of
''shallow water instability'' to be more feasible. 

In Ref.$\cite{Dmitriev1}$, nonlinear evolution of the instability was
considered with account of both the electron fluid viscosity related to
electron-electron scattering and the friction due to electron scattering by
phonons and(or) impurities. The instability has a threshold found for the
case of small viscosity and friction.$^{\cite{Dmitriev1}}$ As expected, the
instability leads, instead of chaotic motion, to the establishment of
low-amplitude stationary oscillations, provided that the threshold current
is exceeded by a small enough margin. In accordance with the theory of weak
turbulence in conventional hydrodynamics$^{\cite{Landau}}$, the amplitude is
proportional to the square root of the current near the instability
threshold. As predicted in Ref.$\cite{Dmitriev1}$, at high enough increments
the instability may result in the formation of step-like electron
concentrations and velocity distributions$^{\cite{Dmitriev2}}$ similar to
shock waves in the conventional hydrodynamics. In Ref.\cite{Rudin1} the
shock wave formation for non-viscous flow in the FET channel for the case of
symmetric boundary conditions was investigated both analytically and
numerically. 

In Ref.$\cite{Cheremisin}$, the instability increment and threshold were
calculated in the special case of strong friction and zero viscosity. The
electron scattering results in the narrowing of an instability region, with
the instability increment decreasing. The instability vanishes at a certain
critical value of friction. The analytical results are in excellent
agreement with computer simulations.

\smallskip Neglecting the friction of the electron fluid, we study the
strongly nonlinear instability mode both analytically and using computer
simulations. First, we check the reliability of our numerical method in the
analytically solvable case of a linear instability mode. Then, we
demonstrate that the instability leads to high-amplitude stationary
oscillations caused by shock wave formation in the FET channel. The shape
and speed of the numerically found shock waves are well described in terms
of the conventional hydrodynamics. Finally, the amplitude of stationary
oscillations is determined.

\section{Analytical approach}

In Ref.\cite{Dyakonov}, it was shown that 2D electrons in a BFET can be
described by the equations analogous to the hydrodynamic equations for
shallow water:

\begin{eqnarray}
\frac{\partial U}{\partial t}+\frac{\partial (VU)}{\partial x} &=&0, 
\eqnum{1}  \label{continuity} \\
\frac{\partial V}{\partial t}+\frac{\partial }{\partial x}\left( \frac{V^{2}%
}{2}+\frac{eU}{m}\right) &=&\frac{\varkappa }{U}\frac{\partial ^{2}V}{%
\partial x^{2}}.  \eqnum{2}  \label{Navier-Stocks}
\end{eqnarray}
Here, the voltage swing, $U=U_{gc}-U_{T}$, corresponds to shallow water
level, $U_{gc}$ is the local gate-to-channel voltage, and $U_{T}$ is the
threshold voltage. Then, $V$ is the local electron flux velocity, $m$ is the
electron effective mass, $\varkappa ${\bf \ }is the viscosity related to
electron-electron scattering. Eq. (\ref{continuity}) is similar to the
continuity equation in hydrodynamics, taking into account the gradual
channel approximation relationship$^{\cite{Shur}}$

\begin{equation}
n_{s}=\frac{CU}{e}.  \eqnum{3}  \label{Shockley}
\end{equation}
Here, $n_{s}$ is the surface electron concentration, $C$ is the capacitance
per unit area. Then, Eq.(\ref{Navier-Stocks}) is analogous to the Euler's
equation in which we neglect the pressure gradient term.$^{\cite{Rudin1}}$
Note, Eqs.(\ref{continuity}-\ref{Shockley}) represent, in fact, the
simplified set of equations. In general, one have to solve the exact
Poisson's and Navier-Stocks hydrodynamic equations. Actually, within gradual
channel approximation the higher order terms in the plasmon dispertion law$^{%
\cite{Eguiliz}}$ as well as the edge-effects contribition to the electric
field are neglected. 

In Ref.\cite{Dyakonov}, it was demonstrated that the steady electron flow is
unstable under the following boundary conditions:

\begin{equation}
U(0,t)=U_{0},\qquad U(l,t)V(l,t)=J/C,  \eqnum{4}  \label{boundary conditions}
\end{equation}
where $U_{0}$ is the fixed source potential($x=0$), and $J$ is the current
per unit gate width, fixed at the drain ($x=l$). In the steady state, $V,U$
are constants: $U=U_{0},V=V_{0}=j/(CU_{0}).$ Neglecting the viscosity, we
can easily solve Eqs.(\ref{continuity}-\ref{boundary conditions}) linearized
with respect to small perturbations of the steady state flow (see Appendix).
As shown in Ref.\cite{Dyakonov}, the instability results in plasma wave
generation, with the FET channel playing the role of a resonator cavity.
Both the eigenfrequencies and the instability increment (see Eq.\ref
{frequency}) depend on $M=V_{0}/S,$ where $M$ is the analog of the Mach
number in hydrodynamics and, $S=(eU_{0}/m)^{1/2}$ is the plasma wave
velocity. The instability exists at $0<M<1$.

We now investigate the strongly nonlinear instability mode, when shock waves
are formed in the FET channel. As well known in the conventional
hydrodynamics, the mass and momentum conservation laws determine the shape
and variables on both sides of the shock wave. With dimensionless coordinate 
$\eta =x/l$, time $\tau =tS/l$, and variables $u=U/U_{0},v=V/S$ introduced,
Eqs.(\ref{continuity},\ref{Navier-Stocks}) yield

\begin{eqnarray}
\frac{\partial u}{\partial \tau }+\frac{\partial (vu)}{\partial \eta } &=&0,
\eqnum{5}  \label{mass conservation} \\
\frac{\partial (vu)}{\partial \tau }+\frac{\partial }{\partial \eta }\left(
uv^{2}+\frac{u^{2}}{2}\right) &=&\kappa \frac{\partial ^{2}v}{\partial \eta
^{2}}.  \eqnum{6}  \label{momentum}
\end{eqnarray}
where $\kappa =\varkappa /lSU_{0}$ is the dimensionless viscosity.

We can easily find a self-consistent solution of Eqs.(\ref{mass conservation}%
,\ref{momentum}) which corresponds to a shock wave propagating in an
infinite medium at fixed velocity, $c$, of the wave front. Simple
integration of Eqs.(\ref{mass conservation},\ref{momentum}) over the
interval containing the shock wave discontinuity yields{\bf \ } 
\begin{eqnarray}
u(v-c) &=&j,  \eqnum{7}  \label{SW equations} \\
vu(v-c)+\frac{u^{2}}{2} &=&\kappa \frac{dv}{d\eta }+q,  \nonumber
\end{eqnarray}
\smallskip where $j,q$ are the current and momentum fluxes which are
continuous in the shock wave. With Eq.(\ref{SW equations}), the shape of the
shock wave can be easily derived as follows:${\bf \smallskip }$

\begin{equation}
\frac{dv_{r}}{d\xi }=v_{r}+\frac{1}{2v_{r}^{2}}+const,  \eqnum{8}
\label{SW shape}
\end{equation}
where $v_{r}=v^{\prime }/j^{1/3}$, $v^{\prime }=v-c$\ is the relative
velocity of the electron flow with respect to the shock wave front, $%
const=(jc-q)/j^{4/3}$. It is noteworthy that for a shock wave propagating in
the infinite medium the dimensionless coordinate, $\xi =\eta j/\kappa
=xjSU_{0}/\varkappa $, does not contain the sample length scale $l$. Then,
according to Eq.(\ref{SW shape}) the spatial distribution of the shock wave
velocity and potential, $u(\xi )=j^{2/3}/v_{r}(\xi )$, are universal
functions of $\xi $. For a low-amplitude shock wave, solving Eq.(\ref{SW
shape}) yields 
\begin{equation}
u(\xi )=\frac{u_{1+}u_{2}}{2}+\frac{u_{2-}u_{1}}{2}\tanh \left( \frac{\xi }{%
\delta }\right) .  \eqnum{9}  \label{SW_low amplitude}
\end{equation}
Here, the indices $1$ and $2$ label, respectively, the values of the
hydrodynamic variables in front of, and behind the shock wave (see Fig.\ref
{Fig.4}a,b), in accordance to Ref.$\cite{Landau}$. Then, $\delta =\frac{%
4u_{1}u_{2}}{3j^{2/3}(u_{2-}u_{1})}$ is the front width of the low-amplitude
shock wave. The smaller the amplitude, the wider the front.

We emphasize that Eq.(\ref{SW equations}) provides that the energy
conservation law is $already$ satisfied. The energy dissipated within the
discontinuity is equal to the difference of energy flux densities $%
j(u+(v^{\prime })^{2}/2)$\ on both sides of the shock wave. Moreover, Eq.(%
\ref{SW equations}) determines the direction in which the variables in the
shock wave change as $u_{2}>u_{1}$, then gives the jump conditions known in
the conventional hydrodynamics.$^{\cite{Landau}}$\ Indeed, for a shock wave
in the infinite medium $\left. \frac{\partial v}{\partial \eta }\right|
_{\pm \infty }=0$, and, therefore the components $u_{1,2}$\ and $v_{1,2}$\
are related by

\begin{equation}
u_{1}=\frac{2v_{1}^{\prime }(v_{2}^{\prime })^{2}}{v_{1}^{\prime
}+v_{2}^{\prime }},\qquad u_{2}=\frac{2(v_{1}^{\prime })^{2}v_{2}^{\prime }}{%
v_{1}^{\prime }+v_{2}^{\prime }},  \eqnum{10}  \label{relationship}
\end{equation}
where $v_{1,2}^{\prime }=v_{1,2}-c$\ is the relative velocity of the
electron flow in front of, and behind the shock wave, respectively.

We now seek the velocities and the potential on both sides of the shock wave
and the speed of the shock wave front, taking into account Eq.(\ref
{relationship}) and the dimensionless boundary conditions $%
u_{1}=1,u_{2}v_{2}=M$ specified by Eq.(\ref{boundary conditions}). In this
case, however, the number of algebraic equations is less than the total
number of variables. Consequently, we set the relative shock wave amplitude, 
$u_{21}=$ $u_{2}/u_{1}$, as a free variable. Finally, the speed of the shock
wave front, $c_{\pm }$ , and the electron flow velocity at the source, $%
v_{1}^{\pm }$ , are given by 
\begin{eqnarray}
c_{\pm } &=&\frac{M}{u_{21}}\pm \upsilon ,  \eqnum{11}  \label{SW speed} \\
v_{1}^{\pm } &=&\frac{M}{u_{21}}\pm \upsilon (1-u_{21}).  \eqnum{12}
\label{source velocity}
\end{eqnarray}
\bigskip Here, the signs correspond to shock wave propagating in the channel
upstream (+) and downstream (-), respectively. Then, $\upsilon =\left( \frac{%
1+u_{21}}{2u_{21}}\right) ^{1/2}$ is the shock wave front speed for a
stagnant electron fluid (i.e. that with $M=0$). It should be noted that the
speed of the low-amplitude ( $u_{21}\simeq 1$) shock wave front coincides
with the plasma wave propagation velocities $M\pm 1$, in agreement with the
conventional hydrodynamics.$^{\text{\cite{Landau}}}$ It will be recalled
that Eqs.(\ref{source velocity},\ref{SW speed}) are valid when $u_{21}>1$.

\section{Computer simulations}

\smallskip Let us first estimate the order of magnitude of the dimensionless
viscosity $\kappa $. In a highly non-ideal 2DEG, with the thermal, Bohr, and
Fermi energies of the same order of magnitude, the viscosity of the electron
fluid is given$^{\cite{Dyakonov}}$ by $V_{F}\lambda _{ee}\sim \hbar /m$,
where $V_{F\text{ }}$ is the Fermi velocity, $\lambda _{ee}\sim n_{s}^{-1/2}$
is the mean free path for electron-electron collisions. For $U_{0}\sim 0.5$%
V, $n_{s}\sim 10^{12}$cm$^{-2}$, and submicrometer( $l=0.2$ $\mu $m )
AlGaAs/InGaAs based FET at $T=77$K we obtain $\kappa =\lambda
_{ee}V_{F}/lS\approx 0.01$. Accordingly, we further assume that $\kappa \ll
1.$

We now compare our analytical results with computer simulation data. The
numerical method used is the well-known Broilovskaya algorithm$^{\cite
{Roache}}$\ adapted for a system with two boundaries. In Ref.$\cite
{Dmitriev2}$, a detailed description of the numerical procedure based on
Eqs.(\ref{mass conservation},\ref{momentum}) was given. As was underlined in
Ref.\cite{Rudin2}, a finite-difference approximation methods always involve
a numerical error which could be associated with a diffusion-like term
similar to that in Eq.(\ref{momentum}). Consequently, we further assume that
within our numerical method there exists a nonzero parasitic ''numerical
viscosity'' $\kappa _{n}$. We will demonstrate that $\kappa _{n}\ll 1$.

In the present paper we use two different numerical approaches. Initially
(case (i)), we omit viscosity term ($\kappa =0$) in Eqs.(\ref{mass
conservation},\ref{momentum}), and then solve them under the boundary
conditions specified by Eq.(\ref{boundary conditions}). We use the first
mode of plasma wave oscillations ( Eqs.(\ref{helix}) at $A=10^{-4},n=1,\tau
=0$) as the initial condition. In Fig.\ref{Fig.1}, the computer simulation
data are presented as the time dependence of the velocity at the source, $%
v(0,\tau )$, in relation to the drain potential $u(1,\tau )$ for different
Mach numbers. In the same figure, we plot the solution given by Eq.(\ref
{helix}), found within the linear approximation. The computer simulation
data are well described by the analytical theory in the linear regime of the
instability evolution for $0<M<1$.{\bf \ }

We now consider the strongly nonlinear instability regime. Instead of a
chaotic motion, as expected, the instability leads to stationary
oscillations irrespective of the initial conditions. The oscillation
amplitude depends on the Mach number. As an example, in Fig.\ref{Fig.2} we
plot four different phases clearly defined in stationary oscillations. In
the first step (Fig.\ref{Fig.2},a), the shock wave appears at the drain and
moves therefrom toward the source. After subsequent reflection from the
source the jump propagates upstream the channel and then disperses (Fig.\ref
{Fig.2},b). The dispersion of the shock wave front is a generic phenomenon
known in the conventional hydrodynamics.$^{\text{\cite{Landau}}}$ In other
phases, there are relatively smooth distributions moving downstream and
upstream the channel.{\bf \ }It is noteworthy that the shallow rapid
oscillations near the discontinuity( see Fig.\ref{Fig.2},a ) are numerical
artifacts that always appear in finite-difference approximation methods for
a non-viscous flow.$^{\text{\cite{Roache}}}$

Let us now examine in more detail the shock wave clearly defined (bold line)
in Fig.\ref{Fig.2},a. It can be seen that a discontinuity approaching the
source can be described analytically in terms of a shock wave in the
infinite medium. To confirm this, we plot in Fig.\ref{Fig.1} the analytical
dependences $v_{1}^{\pm }(M,u_{21})$ given by Eq.(\ref{source velocity}). We
ascertained that the upstream shock wave at the source is well described by
the theory at $0.05<M<1$. It should be noted that the above description
fails to account for the dependence of the amplitude of stationary
oscillations on the Mach number. Nevertheless, we use the computer
simulation data and find the amplitude of stationary oscillations, $%
u_{21}^{st}(M)$, associated with the downstream shock wave (Fig.\ref{Fig.1},
point A). This dependence (see Fig.\ref{Fig.3},a) correlates with that
furnished by Eq.(\ref{frequency}) for the instability increment. The higher
the increment, the stronger the amplitude of stationary oscillations. Using
the dependence $u_{21}^{st}(M)$\ and Eq.(\ref{SW speed}), we can readily
calculate the speed of the upstream and downstream shock waves (see Fig.\ref
{Fig.3}). It can be seen that at $M\rightarrow 0;1$ the speed of the
low-amplitude shock waves coincides with the plasma wave propagation
velocities $M\pm 1$.

\smallskip Up to now, we have used the approach $\kappa =0$, with the
parasitic ''numerical viscosity'' assumed to be small $\kappa _{n}\ll 1$.
Let us examine the opposite situation (case(ii)) $\kappa _{n}\ll $\ $\kappa
\ll 1$\ with viscous term included in Eq.(\ref{momentum}). If this case,
Eqs.(\ref{mass conservation},\ref{momentum}) has higher order and,
consequently, requires an additional boundary condition other than those
given by Eq.(\ref{boundary conditions}). Fortunately, the extra condition
can be readily found with the help of the continuity equation. Indeed, for a
fixed voltage at the source Eq.(\ref{mass conservation}) yields $\left. 
\frac{dv}{d\eta }\right| _{0}=0$. Previously, this condition was derived in
Ref.$\cite{Dmitriev1}$\ on the basis of a more complicated reasoning.{\bf \ }%
With the above set of boundary conditions, we examined numerically the
instability evolution of a viscous flow at $0.005<\kappa <0.05$ and $M<0.15$%
. As expected, in the linear regime of instability, the numerical data
coincide with those found above for the non-viscous case (i). Then, in the
nonlinear regime, the viscosity enhancement results in a minor damping of
stationary oscillations. The amplitude of these oscillations remains nearly
constant in the actual range of $\kappa $, and, hence, we can use the simple
non-viscous approach (i) instead of (ii) in this case.

To conclude, we estimate the parasitic ''numerical viscosity.'' In Fig.\ref
{Fig.4}, we plot the spatial distributions for upstream shock waves(see Fig.%
\ref{Fig.1},a, point A) at fixed Mach number $M=0.1$\ and different
viscosities. it will be recalled that, for a shock wave in the infinite
medium, the spatial distribution $u(\xi )$\ scales with $\xi \sim \kappa $.
Consequently, we can represent the same data re-plotted, e.g. the case $%
\kappa =0.01$. The plots are collapsed(see Fig.\ref{Fig.4},b). In addition,
we plot in the same figure the distribution of the potential in the shock
wave, found from Eq.\ref{SW shape}. It can be seen that the shape of the
numerically found shock wave is well described by the theory. Using the
collapse formalism we can easily estimate the parasitic ''numerical
viscosity'' to be $\kappa _{n}=0.0013$. This value is consistent with
estimations done in Ref.\cite{Rudin2}. To the best of our knowledge, the
exact value of the ''numerical viscosity'' within the Broilovskaya algorithm
is unknown.

In summary, in the nonlinear regime the current instability leads to
high-amplitude stationary oscillations caused by shock waves formation in
the FET channel. The amplitude of stationary oscillations is determined
using computer simulation data. The numerically found shape and speed of the
shock waves are well described in terms of the conventional hydrodynamics.
The parasitic ''numerical viscosity'' is estimated.

\section{Acknowledgments}

The author is grateful to A.P.Dmitriev, G.G.Samsonidze and V.Yu.
Kachorovskii for helpful discussions. This work was supported in part by
RFBR and INTAS, grant YSF 2001/1-0132.

\section{Appendix}

Following Ref.$\cite{Dyakonov}$, we seek the instability by analyzing the
temporal behavior of small perturbations $\delta V,\delta U$ superimposed on
a steady flow, i.e. $V(x,t)=V_{0}+\delta V(x)\exp (-i\omega t),\
U(x,t)=U_{0}+\delta U(x)\exp (-i\omega t)$. Using dimensionless variables
introduced in the text and non-viscous($\kappa =0$) Eqs.(\ref{continuity}-%
\ref{boundary conditions}) linearized with respect to $\delta V,\delta U$ we
obtain 
\begin{eqnarray}
v &=&M+A%
\mathop{\rm Re}%
\left[ \exp (i(k_{1}\eta -\Omega \tau ))+\exp (i(k_{2}\eta -\Omega \tau
))\right]  \eqnum{A-I}  \label{linear solution} \\
u &=&1+A%
\mathop{\rm Re}%
\left[ \exp (i(k_{1}\eta -\Omega \tau ))-\exp (i(k_{2}\eta -\Omega \tau
))\right]  \nonumber
\end{eqnarray}
where $A$ is the arbitrary constant dependent on the initial conditions, and 
$\Omega =\omega l/S$ is the dimensionless frequency. Then, $k_{1,2}=\pm
\Omega /(1\pm M)$ denote the wave vectors for the upstream and downstream
plasma waves respectively. According to Ref.$\cite{Dyakonov}$, both the real
and imaginary parts of the complex frequency $\Omega =\Omega ^{\prime
}+i\Omega ^{\prime \prime }$ of plasma wave generation are given by 
\begin{equation}
\Omega ^{\prime }=\frac{\left( 1-M^{2}\right) }{2}\pi n,\ \Omega ^{\prime
\prime }=\frac{\left( 1-M^{2}\right) }{2}\ln \left| \frac{1+M}{1-M}\right| .
\eqnum{A-II}  \label{frequency}
\end{equation}
where $n$ is an odd integer for $\left| M\right| <1$.

From Eqs.(\ref{linear solution}) we can readily find the time-dependent
velocity at the source $v(0,\tau )$ and drain potential $u(1,\tau )$ as
follows 
\begin{eqnarray}
v(0,\tau ) &=&M+2A%
\mathop{\rm Re}%
\left[ \exp (-i\Omega \tau )\right]  \eqnum{A-III}  \label{helix} \\
u(1,\tau ) &=&1+2A\frac{(1+M)^{\frac{M-1}{2}}}{(1-M)^{\frac{M+1}{2}}}%
\mathop{\rm Re}%
\left[ \exp (i(\Delta \varphi -\Omega \tau ))\right]  \nonumber
\end{eqnarray}
where $\Delta \varphi =\pi n(1-M)/2$ is the phase shift. We emphasize that
Eqs.(\ref{helix}) describe a helix (see Fig.\ref{Fig.1}).

\begin{figure}[tbp]
\caption{Instability evolution diagram for (a) $M=0.1$, (b) $M=0.3$, (c) $%
M=0.5$, and (d) $M=0.7$. The thin line represents the linear regime of
instability, given by Eq.(\ref{helix}). The dashed (dotted) asymptotes are
given by Eq.(\ref{source velocity}) at $u_{21}>1$. The numbers denote the
time slice numbers(see Fig.\ref{Fig.4}). }
\label{Fig.1}
\end{figure}
\begin{figure}[tbp]
\caption{Spatial distribution $u(\xi )$ within the stationary oscillation
period at $M=0.1$. }
\label{Fig.2}
\end{figure}
\begin{figure}[tbp]
\caption{Stationary amplitude $u_{21}^{st}$ and the speed $c_{\pm }$ of the
shock waves at the source(Fig.\ref{Fig.1}, point A) vs the Mach number.}
\label{Fig.3}
\end{figure}
\begin{figure}[tbp]
\caption{The graphic solution of Eq.(\ref{SW shape}): (a) the potential $%
u(\xi )$ and velocity $v_{r}(\xi )$ distributions in the shock wave (b) the
plot $dv_{r}/d\xi $ vs $v_{r}$. Computer simulation data for: (c) spatial
distribution $u(\eta )$ at $M=0.1$ and $\kappa =\kappa _{n},0.005;0.01;0.02.$
and (d) the same curves collapsed with respect to $\kappa =0.01$. Dashed
line: analytical result given by Eq.(\ref{SW shape})}
\label{Fig.4}
\end{figure}

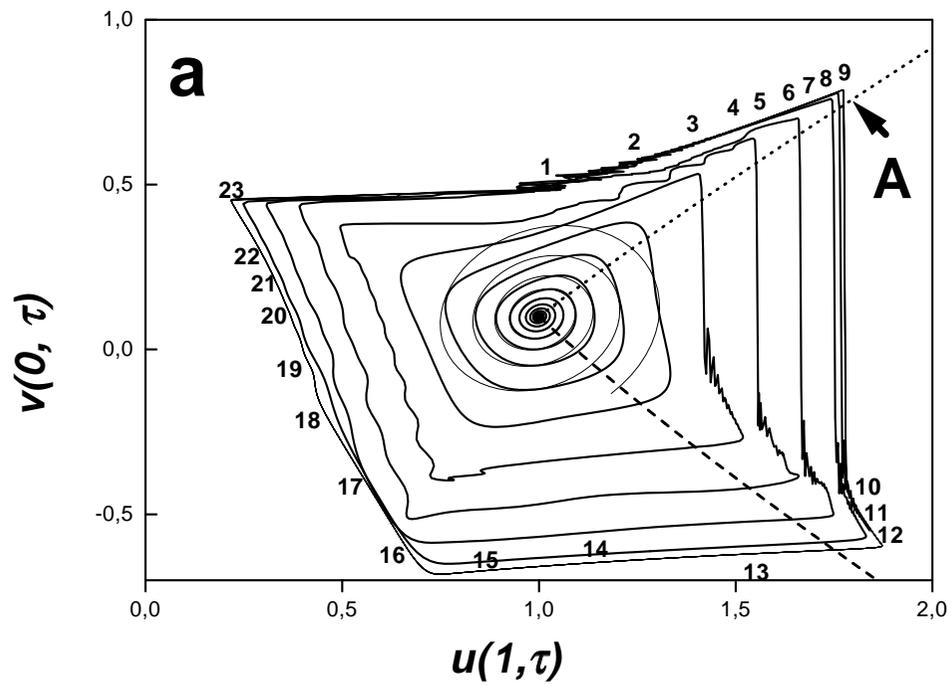
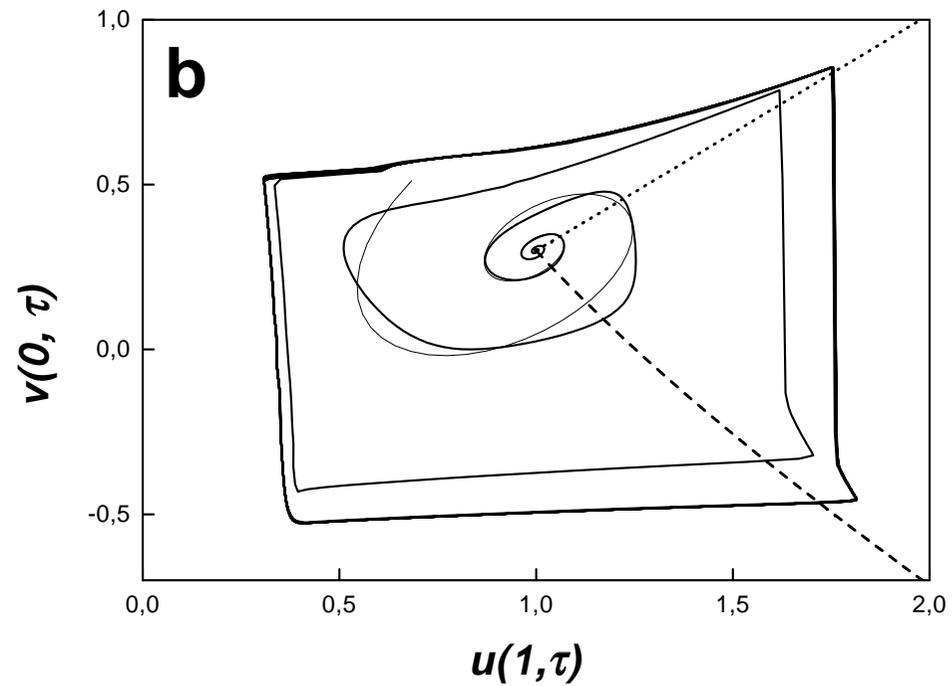
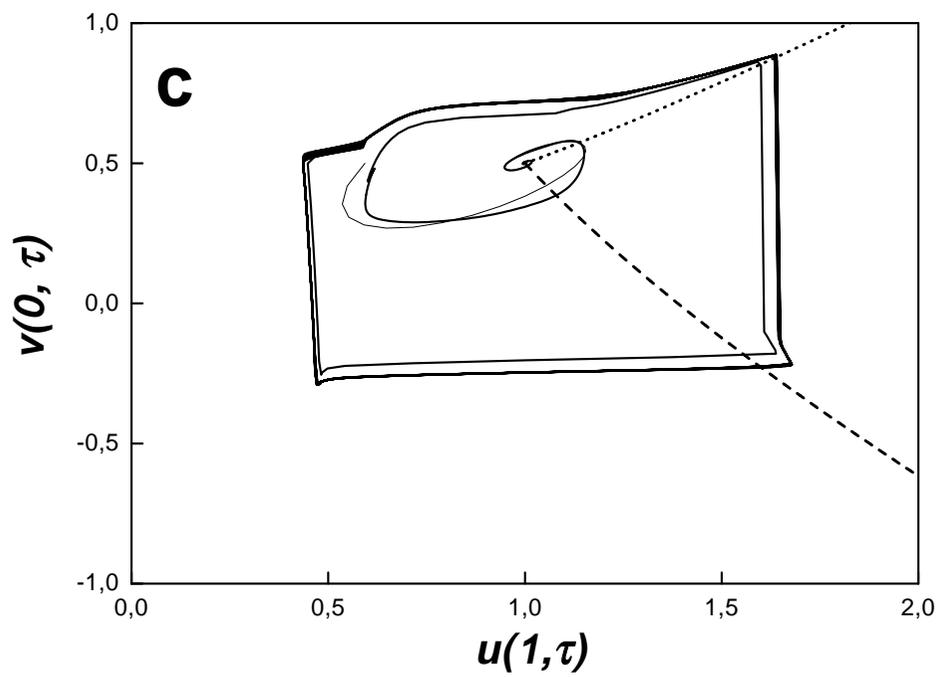
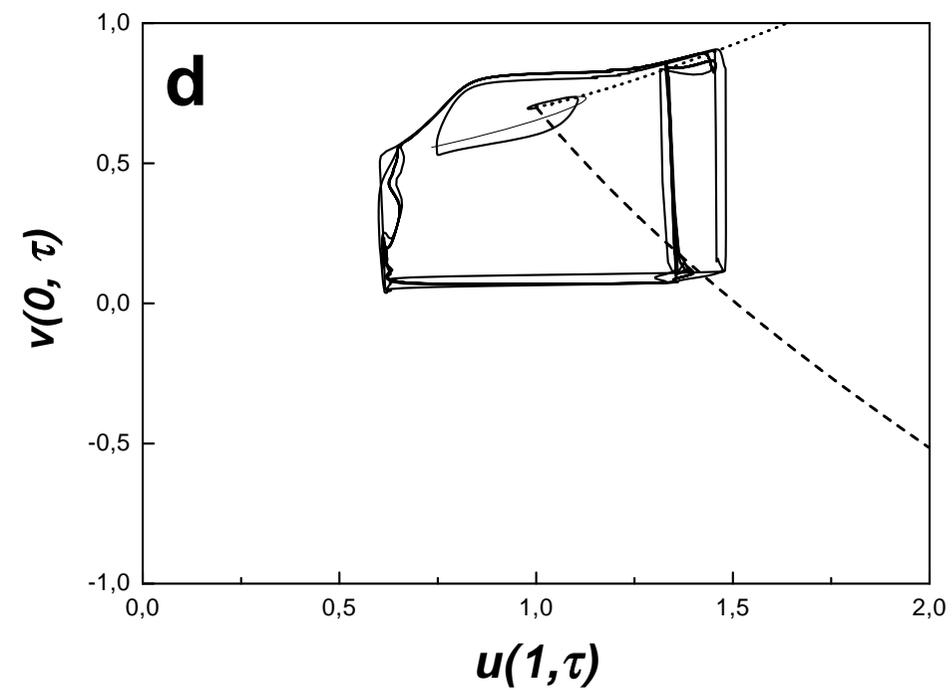
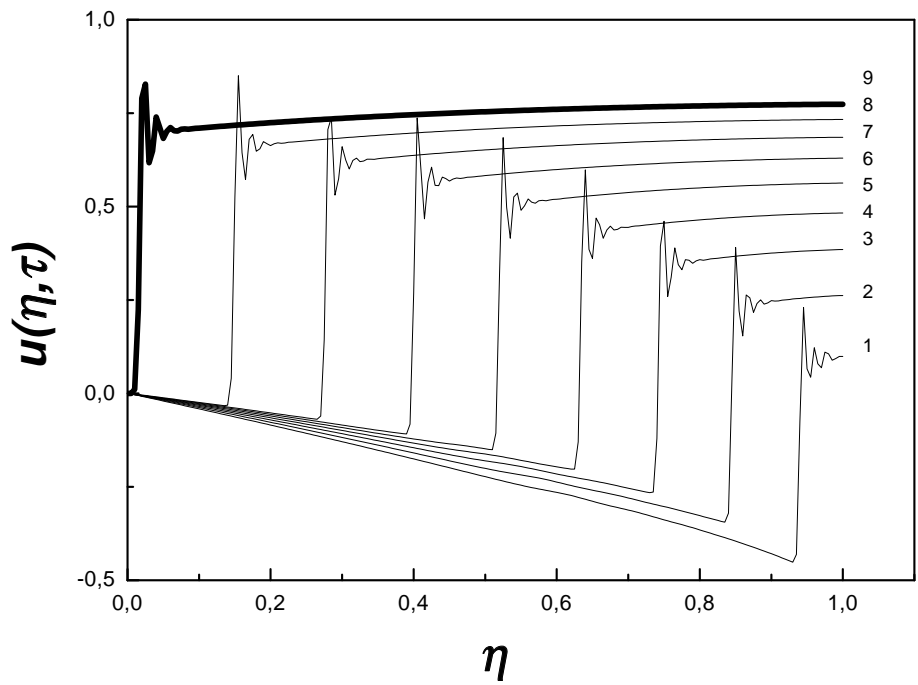
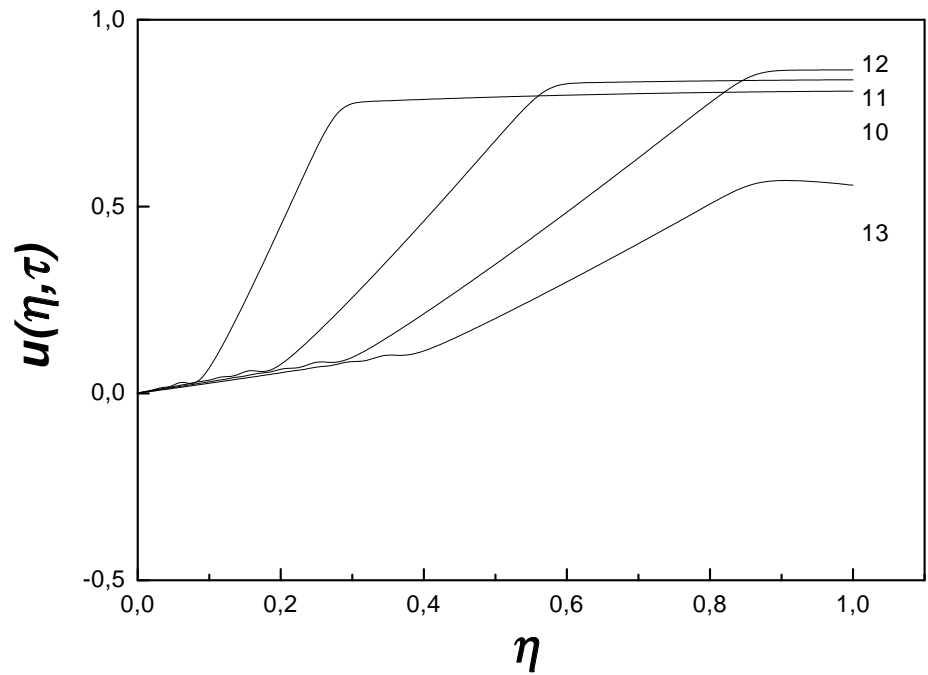
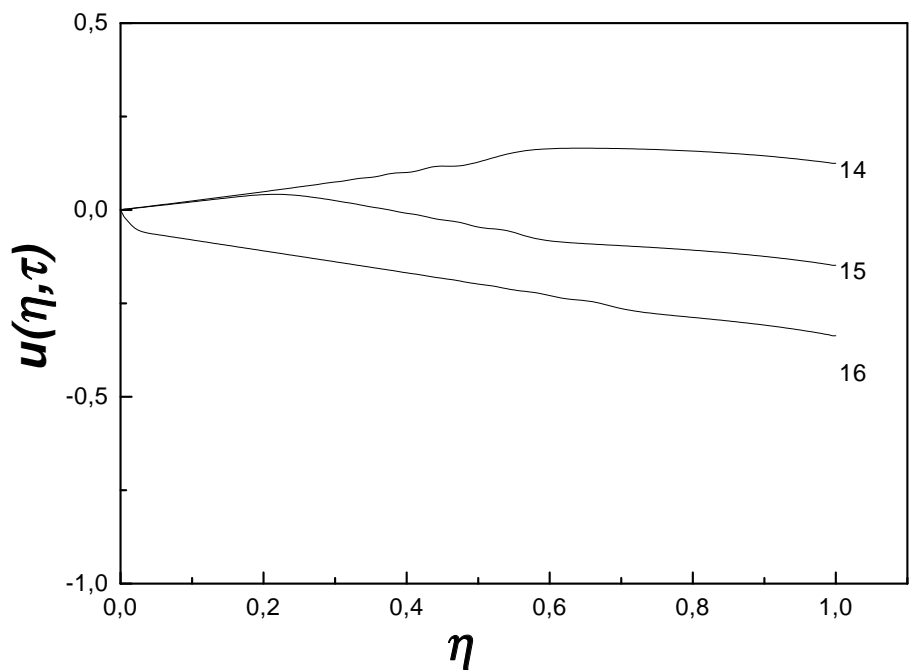
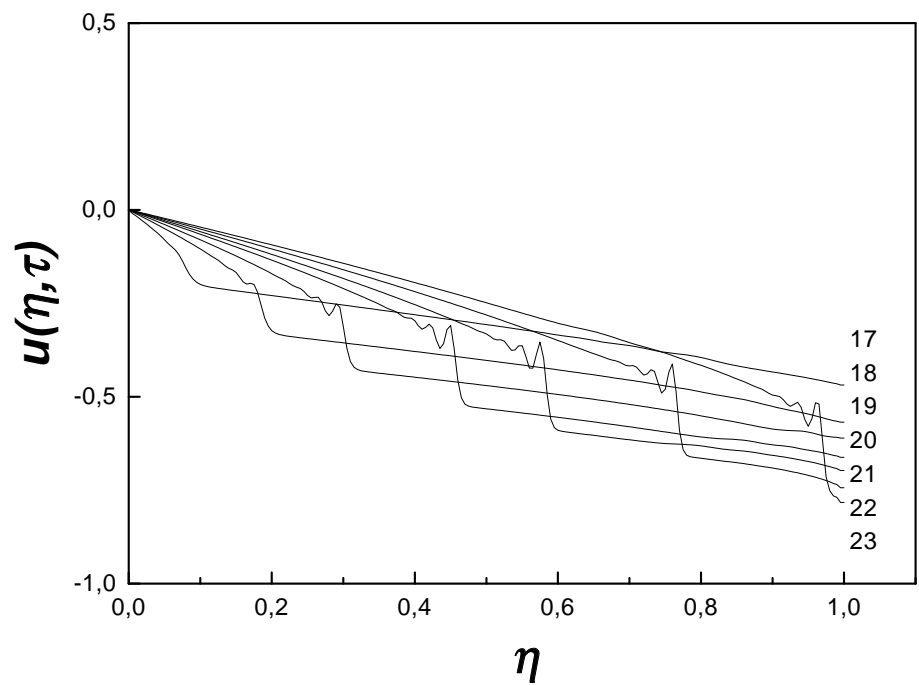
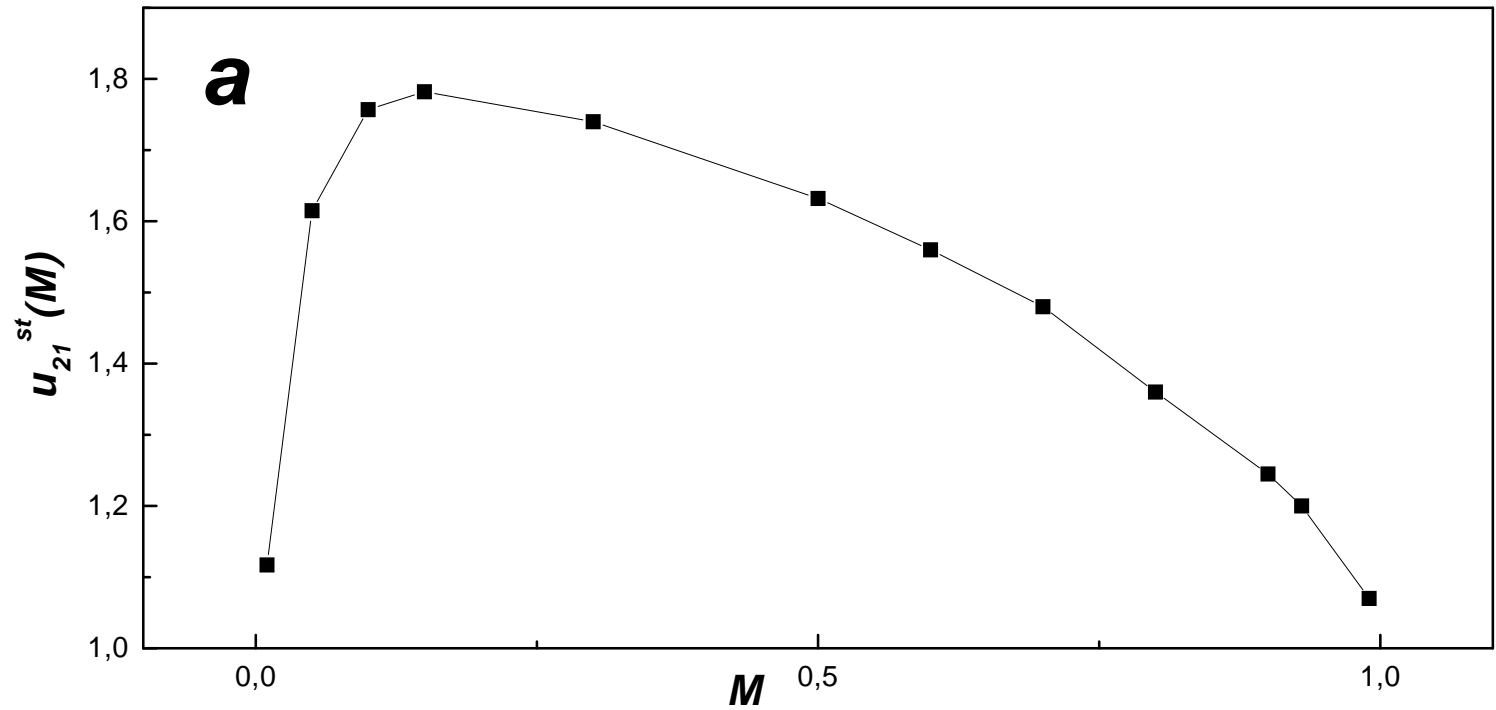
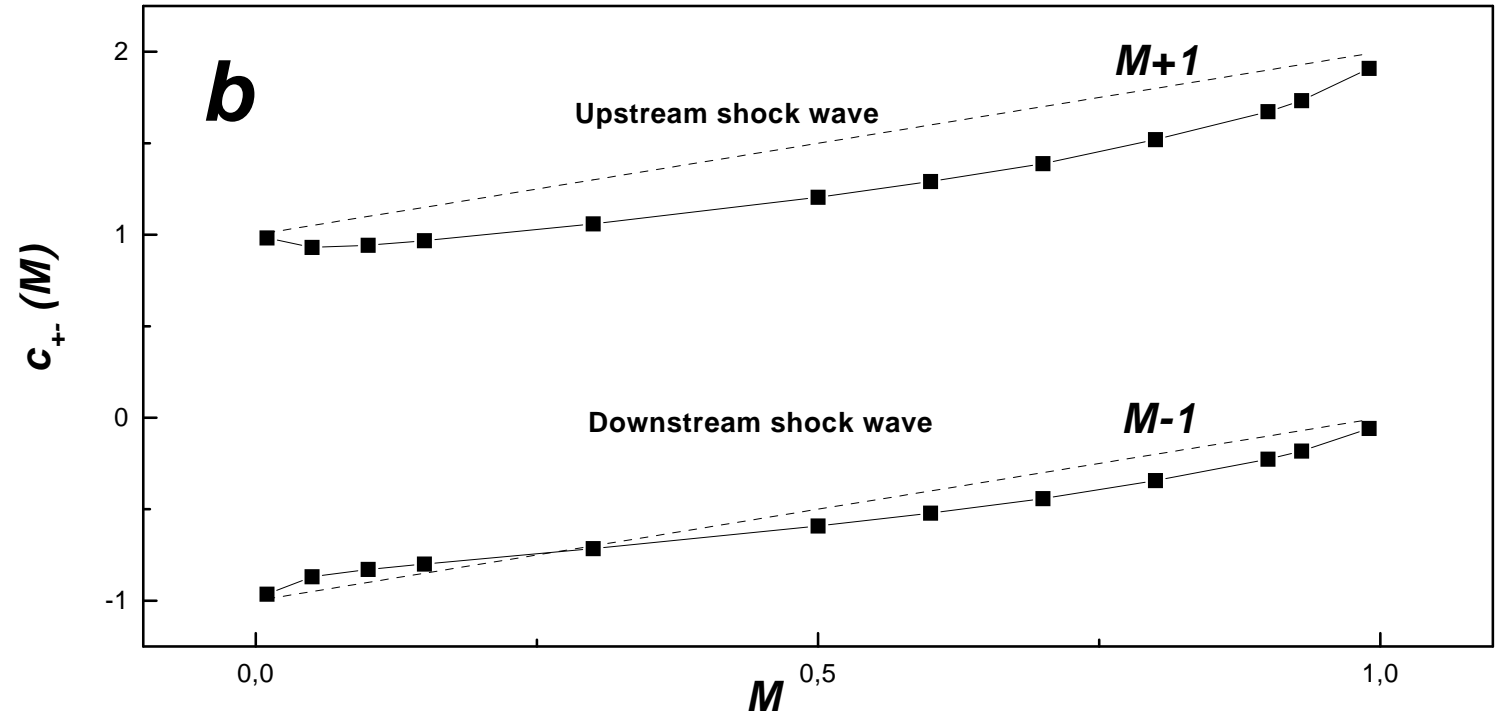
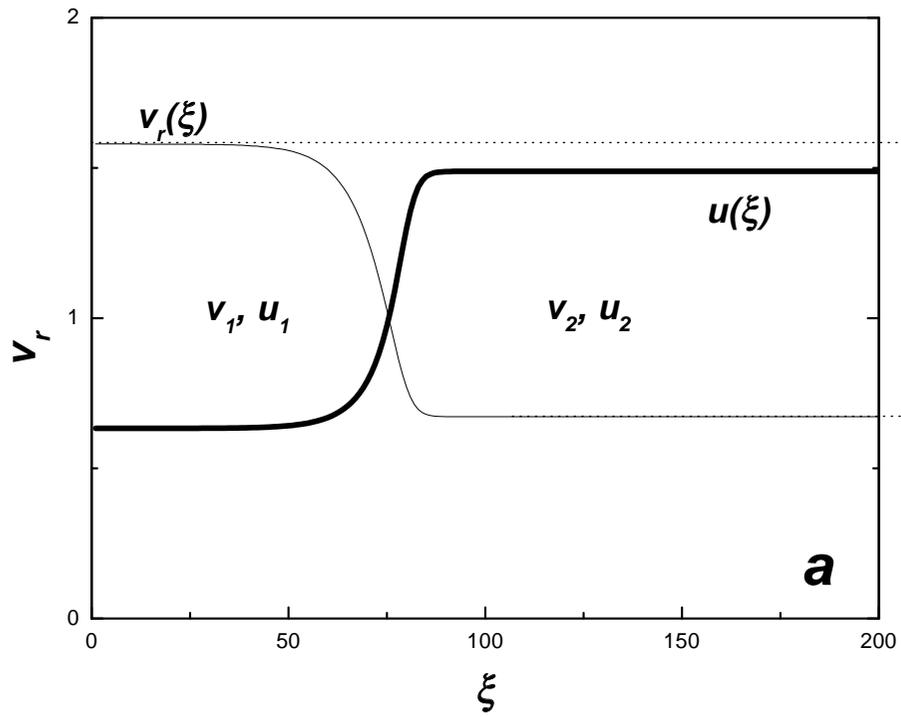
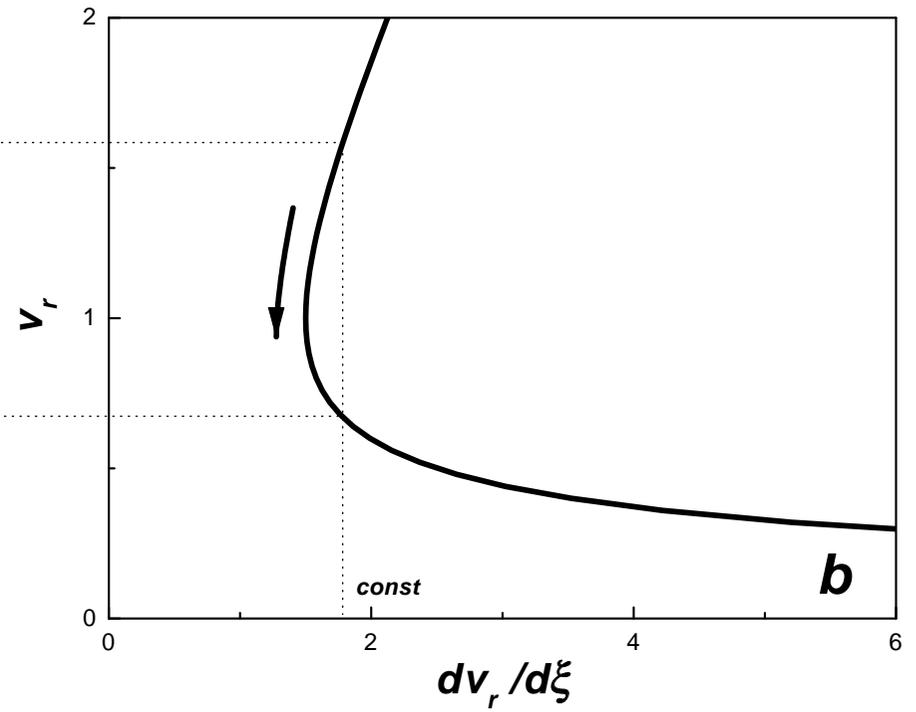
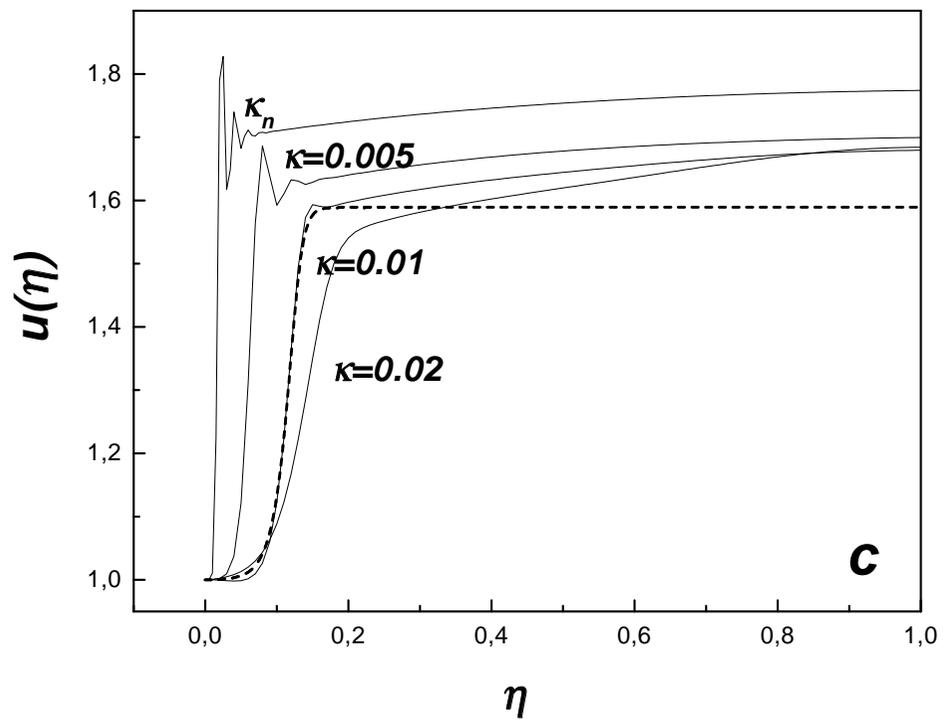
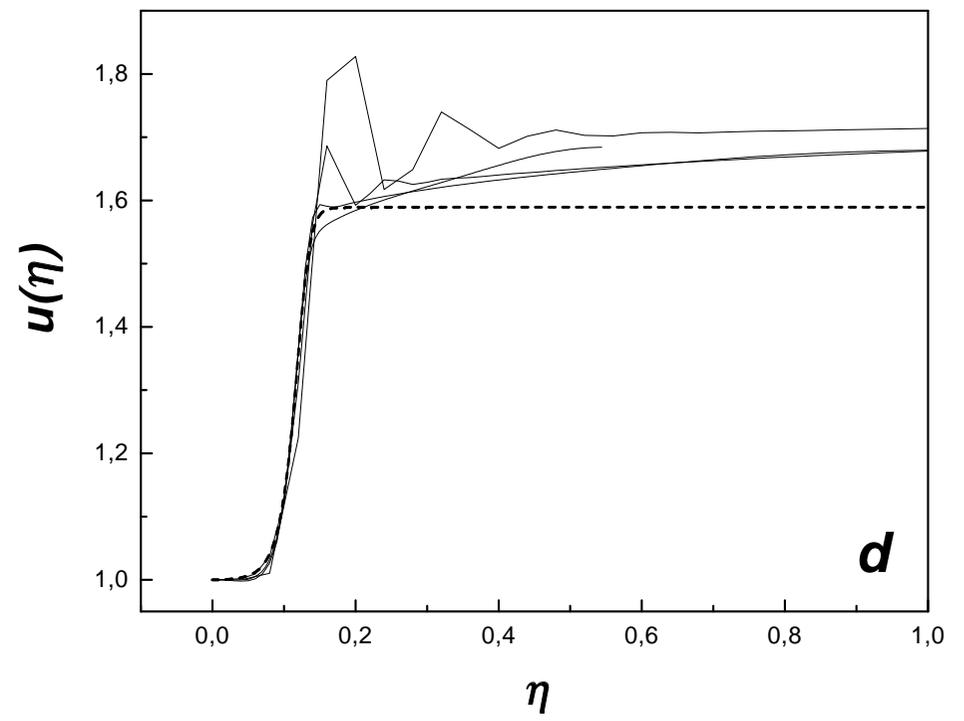
\end{document}